# Ultrafast Microscopy of a Plasmonic Spin Skyrmion

Yanan Dai[1], Zhikang Zhou[1], Atreyie Ghosh[1], Karan Kapoor[1], Maciej Dąbrowski[2], Atsushi Kubo[3], Chen-Bin Huang[4], and Hrvoje Petek[1*]

[1]Department of Physics and Astronomy and Pittsburgh Quantum Institute, University of Pittsburgh, Pittsburgh, PA 15260, USA

[2]Department of Physics and Astronomy, University of Exeter, Exeter EX4 4QL, United Kingdom

[3]Division of Physics, Faculty of Pure and Applied Sciences, University of Tsukuba, 1-1-1 Tenno-dai, Tsukuba-shi, Ibaraki, 305-8571 Japan

[4]Institute of Photonics Technologies, National Tsing Hua University, Hsinchu 30013, Taiwan

## Abstract:

We present an ultrafast microscopy imaging experiment and a general analytical description of a new quasiparticle composed of plasmonic Skyrmion-like spin texture at the core of a surface plasmon polariton (SPP) vortex. The illumination of a circular coupling structure milled in an Ag film by circularly polarized light (CPL) couples its spin angular momentum (SAM) into orbital angular momentum (OAM) of SPPs launching them to form a plasmonic vortex. The coupling of the cycloidal motion of the SPP polarization at the 2D interface, with their orbital swirl at the vortex core causes the plasmonic field to generate 3D SAM pseudovectors, whose topological texture has integer Skyrmion number and is homotopic to a twisted magnetic Skyrmion quasiparticle with the boundary defined by an optical L-line singularity contour. An analytical description finds that the dielectric discontinuity at the Ag/vacuum interface supports on each side entwined twisted Skyrmion pairs that are characterized by stable topological textures with opposite Skyrmion numbers. The SAM texture of the Skyrmion pair within the primary vortex ring corresponds to a monopole-hedgehog type SAM texture, with a SAM singularity at the vortex core. Interferometric time-resolved two-photon photoemission electron microscopy (ITR-2P-PEEM) imaging of the nanofemto spatiotemporal evolution of the SPP fields and simulation by an analytical model, establish the twisted topological plasmonic SAM Skyrmion quasiparticle at the vortex core. The SAM textures can probe and simulate topological responses in trivial and topological materials that can be coupled in the near-field of the SPP vortex. The theory anticipates different field structures and the accompanying topological spin textures that construct single Skyrmion and meron-like quasiparticles as well as their arrays.

## I. Introduction

Topological defects pervade phase transitions in physics, from the context of structure-formation in cosmology, to quantum fluids in condensed matter systems [1-4]. Skyrmions, the archetypes of such defects, were originally proposed as a soliton-like solutions of nonlinear field theory for the structure of nucleons [5], and later found in condensed mater contexts as homotopic stable textures of spins in ferromagnetic materials and Bose-Einstein condensates, molecular orientational order in liquid crystals, and fluxon pinning by superconducting vortices [6-9]. In magnetic materials, they can be stable structures within ferromagnetic phase diagrams where spin-spin interaction described by the Dzyaloshinskii-Moriya (DM) Hamiltonian favors the spin Skyrmion texture formation, which can be experimentally imaged in real space by Lorentz microscopy [10], soft X-ray microscopy [11], spin-polarized scanning tunneling microscopy [12], spin-polarized low energy electron microscopy [13], etc., or indirectly deduced in the reciprocal space by X-ray diffraction [14]. The topologically stable magnetic Skyrmions can be created and annihilated by spin polarized currents, and be transported by low in-plane electric currents, making them attractive as bits of magnetically stored data or in other quantum spintronic devices [11, 15]. While Skyrmions are characterized by integer Skyrmion numbers, *i.e.* topological charge of the texture, a related topological quasiparticle, meron, which is characterized by half-integer Skyrmion number, forms similar magnetic spin textures through the DM interaction; we have imaged the plasmonic meron quasiparticle generation and stability by ultrafast microscopy and described by analytical theory [16].

Skyrmion textures are characterized by stereographic projection of 3D spin pseudovectors onto 2D bisecting surfaces with spin directions encompassing the 4π solid angle of a sphere [10-12, 14, 17]. Magnetic Skyrmions are dominantly classified into two limiting categories: the Néel type Skyrmion where the spin texture rotates cycloidally in the plane containing its rotational symmetry axis; and the Bloch type Skyrmion, whose spins rotate in a plane perpendicular to the rotational symmetry axis [14, 18, 19]. It has been reported, however, that twisted Skyrmions can also exist in bulk magnetic materials, where the spin texture is intermediate between the Néel or Bloch limits. These different textures provide understanding of the static stabilization of topological defects in different phases through the proximate vector interactions, and can be considered as an additional degrees-of-freedom for the structure formation and potentially for information storage and retrieval [14, 18].

Whereas magnetic Skyrmions are stable quasiparticles that can be characterized by standard microscopic methods, their creation and annihilation dynamics are not readily accessed by experiment. The ultrafast dynamical creation and imaging of Skyrmion-like textures in optics, especially within 2D surface electromagnetic modes at metal/dielectric interface such as SPPs [20, 21], opens the possibility of resolving the dynamics of the formation and interactions of Skyrmions among themselves and with other materials. Skyrmion-like plasmonic field and SAM textures that can be generated by optical excitation of textured metal surfaces have been reported [16, 22, 23]. Therefore, by employing monochromatic continuous wave (CW) to femtosecond light sources [24-26], they could potentially be generated and function as information bits or active elements in ultrafast information storage and processing nodes [27, 28], where their phases, topological spin textures, and coherent interactions with other materials could enable applications such as quantum computing [29], or investigations of correlated dynamical processes such as topological phase transitions [19, 30]. In addition, the Skyrmion-like spin textures generated by plasmonic fields can be designed to enhance quasiparticle interactions, such as plasmon-exciton and plasmon-phonon coupling in, for example, valleytronic van der Waals semiconductors [31-35], with potential applications to optoelectronic and photocatalytic functions, with a reduced size and increased capability [36-38], thus expanding the diversity in the field of topological photonics [39] Such quasiparticles can be created and dissipate on femtosecond time scales defined by the optical generation pulses, and can be investigated in *dynamical* contexts where the Skyrmion-like topological textures are created, and persist within duration of the optical fields or polarization waves in condensed matter down to a few femtosecond time scales, and interact among themselves or other optically coupled materials [16].

SPPs are coherent electromagnetic waves at metal/dielectric interfaces that are supported by a change in sign in the real parts of dielectric functions of the component media; structuring their phase fronts in both space and time [29] can be used to realize complex electromagnetic phenomena such as plasmonic vortices [40, 41], or design of plasmonic devices based on the plasmonic spin-Hall effect [42-47]. The remarkable topological properties of SPPs can be ascribed to their chiral $k_{spp}$ -vector-transverse SAM locking, which is a consequence of their cycloidal field evolution: As SPPs propagate, within single oscillation, their E field cycles from transverse to longitudinal polarization, which locks their SAM in the transverse direction [20, 48, 49]. The plasmonic spin-Hall effect describes the SPP chirality: reversing their $k_{spp}$ k-vector causes their SAM to change sign, just like it happens for circularly polarized light. This key feature enables SPPs to mold topological spin textures.

Furthermore, the topological SAM textures are defined by the geometry of the coupling structures and the generating light polarization. We develop a general analytical model that can be applied to the generation and dynamics of plasmonic topological spin textures. Based on the Huygens principle of wave superposition, we describe how the geometrical phase, in the present case provided by the helicity of a circular metallic coupling structure interacting with circularly polarized light, leads to vortex formation from the defined SPP phase front. The phase fronts of the generated SPP fields are designed to convert the longitudinal SAM of light, which is constrained to $\sigma\hbar$, where $\sigma=\pm 1$, into a plasmonic vortex defined by OAM $L=\pm 1$ [46]. As we will show, once propagating SPPs converge into a vortex, their transverse SAM in the vacuum side of the interface forms a twisted Skyrmion-like texture, which is homologous to a magnetic twisted SAM Skyrmion, and hereafter we will refer to it just as Skyrmion. We characterize the topological character of the Skyrmion up to its boundary at an L-line optical singularity [50, 51], where the superposition of counterrotating in-plane SPP circular polarizations superpose to form a linear polarization boundary, and find that its texture is intermediate between a Néel and Bloch type with an integer Skyrmion Number *N*=±1. In addition to the SAM texture projecting into the dielectric phase of the interface, a complimentary and inseparable mirror Skyrmion forms within the metal with the opposite Skyrmion number.

To resolve with nanometer and femtosecond resolution the ultrafast creation, evolution, and decay of the plasmonic SAM Skyrmion, we employ the interferometric time-resolved two-photon photoemission electron microscopy (ITR-2P-PEEM). This ultrafast microscopy method directly maps the SPP vortex field by recording the distribution of photoelectrons emitted from sample surfaces through the coherent annihilations of two energy quanta from either the plasmon, the photon, or both fields [41, 52]. We apply the ITR-2P-PEEM ultrafast microscopy to record the amplitude and phase of the local surface fields [40, 52-57], from which we can deduce the transient formation of a plasmonic SAM Skyrmion texture based on Maxwell's equations [58]. The derived plasmonic SAM distribution, consistent with our analytical model, directly defines the vectorial spin texture at the core of the plasmonic vortex on the femtosecond time scale, thus bringing an ultrafast picture of the SAM textures in the fundamental plasmonic vortex mode ($L = \pm 1$). The observed plasmonic SAM Skyrmions can be transiently created and annihilated on the time scale of the optical generation pulse and be structured on sub-wavelength spatial scale. Within this nanofemto regime, they can interact with other condensed matter particles and quasiparticles, such as steering conduction electrons in 2D optoelectronic circuits and by resonant energy transfer through spin-specific valley excitons in van der Waals materials [35]. They can also assist ultrafast chemical reactions and enhance dipole interactions with molecules through coupling with the plasmonic fields and spins [59-62]. Because their topological textures break the time-inversion symmetry, plasmonic Skyrmions may find applications in coherent opto-magnetic devices based on the inverse Faraday effect [63, 64], in generation of charged particle vortices [65, 66], and for quantum information processing [67, 68].

## II. Analytic Theory of Plasmonic Skyrmions
## A. Plasmonic Vortex

SPPs are interfacial electromagnetic fields at metal/dielectric interfaces. Their propagation $k_{spp}$ vector is complex because the charge density fluctuations are damped and normal to the interface their wave vector, $\kappa$, is imaginary because their field is evanescent [20, 21]. Their transverse SAM $S$ is given by [20, 69],

$$\boldsymbol{S} \sim \frac{1}{4\omega}\text{Im}[\varepsilon(\boldsymbol{E}^* \times \boldsymbol{E}) + \mu(\boldsymbol{H}^* \times \boldsymbol{H})] \qquad (1)$$

where $\omega$ is the oscillation frequency of the electric $\boldsymbol{E}$ and magnetic $\boldsymbol{H}$ field components of the SPP wave, and $\mu$ and $\varepsilon$ are the permeability and permittivity of the supporting materials (vacuum dielectric or Ag film; although interfacial dielectric functions are nonlocal [70], for the purpose of mathematical description of SPPs we assume that they are local and given by their bulk values); the asterisks indicate complex conjugates. When circularly polarized light (CPL) with helicity σ illuminates a circular coupling structure (Fig. 1a), SPP waves are launched from the structure such that their transverse (out-of-plane, $E_z$) component accumulates 2π phase along the structure circumference to create a wave with OAM. The $E_z$ field converges to a phase singularity at the vortex core [45, 46]. Therefore, a plasmonic vortex is generated with a light helicity dependent OAM of *L*=σ, which can exert optical force on nanoscale objects, causing nanoparticle rotation [71], or drive electron currents [65, 66, 72, 73].

Figure 1(a) defines the coupling structure and the cylindrical coordinate system, which we use to excite and describe the SPP fields that form the plasmonic vortex and its Skyrmion SAM texture. The coupling structure in experiment is a circular slit fabricated by focused ion beam milling of an Ag film [16]. The coordinates that describe the SPP field generation and propagation into a vortex are $r$ and $\theta$, while the coupling structure is defined by its radius $R$ and azimuthal angle $\emptyset$. The circularly polarized excitation light propagates in the -*z*-direction to interact with the coupling structure; its *E*-field is described by $\boldsymbol{E}_{ex} = \frac{1}{\sqrt{2}}e^{\pm i\emptyset}(\widehat{\boldsymbol{R}} + i\widehat{\emptyset})e^{i(k_L \cdot z - \omega_L t)}$, where $\widehat{\boldsymbol{R}}$ and $\widehat{\emptyset}$ are the radial and tangential unit vectors, and $\boldsymbol{k}_L$ and $\omega_L$ are the wave vector and frequency of light, respectively. For an infinitesimally thin radial slit coupling structure, however, only the radial in-plane circulating component of the optical field will couple into the SPPs [56]. The wavelength of excitation in experiment and simulations is $\lambda_L$=550 nm, corresponding to the wavelength of SPPs of $\lambda_{spp}$=530 nm; the model, however, is universal for any excitation wavelength or material that supports SPPs.

Often, plasmonic vortices are discussed only in terms of the transverse component of the SPP fields [74], because the ratio between the field amplitudes of the transverse and longitudinal components is $\sim\sqrt{\varepsilon_m/\varepsilon_d}$, where $\varepsilon_{m(d)}$ is the dielectric function of metal (dielectric). Specifically, the transverse component of SPP field introduced by a coupling structure at radius $R$ can be described by $E_{0z}e^{i(\emptyset-\omega_{spp}t)}e^{-\kappa z}Rd\emptyset\hat{\mathbf{z}}$, where $E_{0z}$ is the field amplitude, $e^{i(\emptyset-\omega_{spp}t)}$ is the time dependent phase of the SPP fields of frequency $\omega_{spp} = \omega_L$, $e^{-\kappa z}$ is the exponential decay of the evanescent field amplitude normal to the interface, and $Rd\emptyset$ is the infinitesimal angular increment along the coupling structure circumference. By the Huygens principle, the coupling structure can be considered as an array of point sources along its circumference that launch SPP waves when exposed to $\boldsymbol{E}_{ex}$. Therefore, one can obtain the SPP fields near the vortex core by propagating all contributions emanating from the point sources along the coupling structure with the appropriate phase and integrating them.

In latter discussions, we only consider the limiting $z$=0 fields from the vacuum side, because we are interested in them and SAM at the interface. In addition, while calculating the transverse components near the vortex core requires only a scalar integration along the $z$-direction, the integrand for obtaining the longitudinal components needs a vector projection along the radial and tangential axes (this introduces a dot product in the integrand; supplementary material).

For the $L$=-1 plasmonic vortex, the field components on the vacuum side can be expressed by eqs. (2), where $0^+$ indicates the dielectric side, $\varepsilon_0$ ($\varepsilon_d$) is the absolute (relative) permittivity of the vacuum, and $J_\nu$ is a Bessel function of the first kind of order $\nu$, which describes the areal distribution of SPP in the interface plane (the detailed derivations of eqs. (2) are provided in supplementary material).

$$E_z(r,\theta,0^+) = i\frac{2\pi R k_{spp}}{\omega_{spp}\varepsilon_0\varepsilon_d}e^{ik_{spp}R}e^{i\theta}J_1(k_{spp}r) \qquad (2a)$$

$$E_r(r,\theta,0^+) = -\frac{\pi i \kappa}{\omega_{spp}\varepsilon_0\varepsilon_d} e^{ik_{spp}R} e^{i\theta}\left[2irJ_1(k_{spp}r) + R\left(J_0(k_{spp}r) - J_2(k_{spp}r)\right)\right] \quad (2b)$$

$$E_\theta(r,\theta,0^+) = \frac{\pi \kappa R}{\omega_{spp}\varepsilon_0\varepsilon_d} e^{ik_{spp}R} e^{i\theta}\left[J_0(k_{spp}r) + J_2(k_{spp}r)\right]. \quad (2c)$$

The distribution of $|E_z|$ from eq. (2a), which is the vortex field carrying a OAM of $L$=-1 (left circularly polarized σ=-1), is plotted in Figure 1(b) for $R = 2\lambda_{spp}$; it shows that $E_z$ is maximized at the first peak of the Bessel function, $J_1(k_{spp}r)$, at $k_{spp}r_1 \approx 1.84$. From this we establish that the radius of the primary vortex ring, where the SPP field is the strongest and therefore the 2PP signal is the most intense, occurs at $r_1 \approx 0.29\lambda_{spp}$ (dashed circle). The superimposed arrows in Figure 1(b) indicate the Poynting vectors, which show the counterclockwise SPP flow where it forms the vortex. Because of the transverse SAM is locked to the counterclockwise $k_{spp}$, its association with the plasmon vortex can be obtained based on the right hand rule: at the vortex core it should point into the dielectric, and from there diverge to lie in the surface plane and point away from the core at the primary vortex ring [Figure 2(b); henceforth, the vortex ring]. By reversing the light helicity σ, we obtain an $L$=+1 vortex with the opposite circulation, where the transverse SAM at the vortex ring points towards the core, and at the core, into the metal.

For the TR-2P-PEEM imaging, we also need to evaluate the longitudinal (in-plane) component of the SPP fields ($\boldsymbol{E}_\parallel = \boldsymbol{E}_r + \boldsymbol{E}_\theta$); this component is often overlooked because it has a lower amplitude than the out-of-plane component, but it also has a spin angular momentum, which must be considered to describe the overall SAM topology. For an $L$=-1 SPP vortex, the radial ($|E_r|$) and tangential ($|E_\theta|$) electric field components are plotted in Figure 1(c)(d). The $|E_r|$ component has a focus at the vortex core, unlike $|E_z|$, which creates the phase singularity. Moreover, $|E_r|$ reaches the minimum at $r_1$ (vortex ring), indicating that there the SPP fields undergo purely tangential orbiting motion.

Additionally, the $|E_\theta|$ component is also focused at the vortex core, but its amplitude decays much more slowly than $|E_r|$. This non-zero tangential component is related to the tangential $k_{spp}$-vector, which is responsible for the counterclockwise field evolution near the vortex. The minimum of $|E_\theta|$ occurs at $r_2 \approx 0.61\lambda_{spp}$, which is a zero of $J_1(k_{spp}r)$. While Figure 1 shows a static snapshot of the time-integrated plasmon fields, their dynamical properties are determined by the Poynting vectors as in Figure 1(b), which illustrate that the combined field components undergo a counterclockwise orbiting motion around the vortex core, as shown in Figure S1. Therefore, based on the right hand rule, the counterclockwise orbiting motion of $E_r$, which has the same field evolution as the electric field in a left circularly polarized light, produces an out-of-plane SAM at the vortex core.

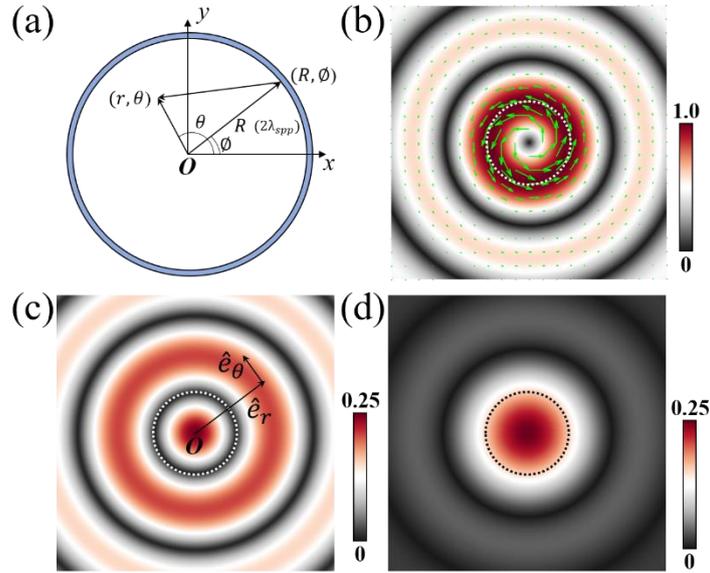

**Figure 1. (a) Schematic of the SPPs ring coupling structure with $R = 2\lambda_{spp}$ (all images) and the definition of coordinates. (b) The $|E_z|$ component of the SPP field (color scale) generated by illumination of the coupling structure with left circularly polarized light, (σ =-1) to form a vortex field distribution, which carries OAM L=-1. The superimposed arrows indicate the Poynting vectors, and the dashed circle (all images in Fig. 1, Fig. 2 and Fig. 7) is the first maximum of $E_z$, which defines $r_1$, the primary vortex ring. (c) The $|E_r|$ field distribution, which is tightly focused at the vortex core, and has the first minimum at $r_1$; this distribution occurs because it is $\frac{\pi}{2}$-phase shifted with respect to $E_z$. (d) The $|E_\theta|$ distribution, which is focused at the vortex core and slowly decays outwards. The color scales of (c) and (d) are ¼ of (b).**

## B. Plasmonic SAM Skyrmion

Having defined the vectorial SPP fields of an *L*=-1 vortex in eqs. (2), we next derive the spatial distribution of their SAM based on the fields in eq. (1) (see supplementary material). The resulting spatial distribution of the SAM components on the vacuum side of the interface are given in eqs. (3):

$$S_r(r,\theta,0^+) = \frac{8\pi^2 R^2 \kappa}{\omega_{spp}{}^2 \varepsilon_0 \varepsilon_d} \frac{J_1^2(k_{spp}r)}{r} \tag{3a}$$

$$S_\theta(r,\theta,0^+) = -\frac{8\pi^2 R r k_{spp} \kappa}{\omega_{spp}{}^2 \varepsilon_0 \varepsilon_d} J_1^2(k_{spp}r) \tag{3b}$$

$$S_z(r,\theta,0^+) = 2\pi^2 R^2 \left(\frac{\kappa^2}{\omega_{spp}{}^2 \varepsilon_0 \varepsilon_d} + \mu_0\right) [J_0^2(k_{spp}r) - J_2^2(k_{spp}r)]. \tag{3c}$$

The $S_z$ component in the metal side ($0^-$) has an equivalent form, but with the appropriate dielectric function ($\varepsilon_d \to \varepsilon_m$), which reverses the sign of $S_z$. Whereas the sign of the in-plane SAM is preserved, because the both handedness of the transverse SPP fields and the real part of the dielectric functions, have opposite sign above and below the interface. This unique texture across the metal/dielectric interface will be detailed in later discussion.

Based on eqs 3, we plot the corresponding normalized 3D SAM texture within the vacuum in Figure 2(a), where one can see that it resembles that of a twisted magnetic Skyrmion. In this texture, SAM reorients from pointing up at the core to pointing down at its periphery, which will be defined [14, 18]. The continuous reorientation of normalized SAM is more clearly illustrated in Figure 2(b), where at a distance $r_1$, $S_{z,n} = 0$ according to the recurrence relations of Bessel functions. Proceeding to a larger *r*, $S_z$ becomes negative, and completes a full $2\pi$ rotation at $r_2$, which we define as the Skyrmion boundary because the in-plane SAM vanishes there. The in-plane component of the normalized SAM, $\mathbf{S}_{\parallel,n} = \mathbf{S}_{r,n} + \mathbf{S}_{\theta,n}$, which is represented by arrows in Figure 2(b) and is described by $J_1^2(k_{spp}r)$, is $\frac{\pi}{2}$ phase shifted with respect to $S_z$ because its amplitude changes from zero at the vortex core, to the maximum at $r_1$ where $\frac{dJ_1(x)}{dx}|_{r_1} = 0$, and back to zero at $r_2$. This demonstrates an interchange between the in-plane and out-of-plane SAM. The distribution of the in-plane SAM $|S_{\parallel,n}|$ is shown in Figure 2(c), where one sees that its maximum is located at $r_1$, which is connected to the maximum of $E_z$; the radial component of the SAM, by contrast, is generated by the cycloidal fields involving $E_z$ and $E_\theta$, which are perpendicular and parallel to the surface. Note that there is also a non-zero tangential component of the SAM, which is proportional to the dimension of the vortex through $r$; it appears because $r$ is not treated as a negligible quantity with respect to $R$ in deriving the Bessel function orders.

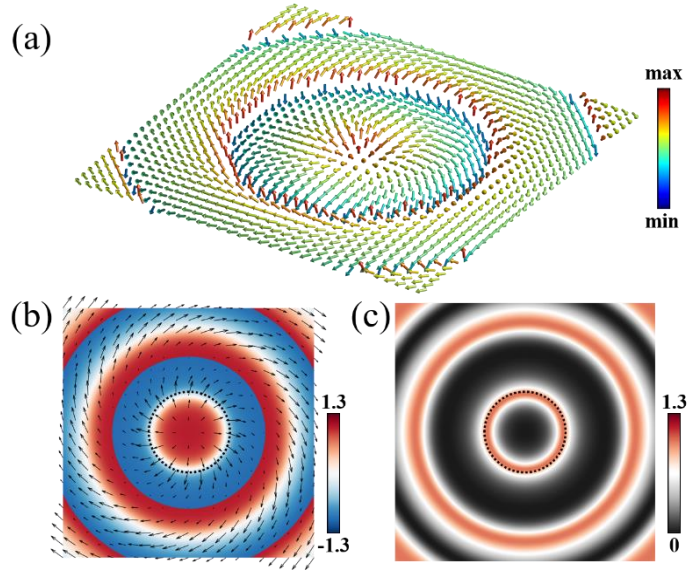

**Figure 2. (a) 3D representation of SAM of the plasmonic Skyrmion. (b) $S_{z,n}$ distribution at the vortex core (color scale). The arrows indicate the directions and amplitudes of the in-plane SAM, $S_{\parallel,n}$. (c) Amplitude distribution of $S_{\parallel,n}$, which is maximized where the $E_z$ field is minimized. Dashed circles indicate vortex ring at $r_1$.**

Because of the finite $S_\theta$ component of SAM, for coupling structures where $r \sim R$, the $S_\theta$ component twists the SAM texture away from a Néel type. To quantify the twist due to $S_\theta$, we introduce the helicity angle $\chi$, which is used to describe magnetic Skyrmions, where it quantifies the angle of the in-plane spin orientation [14, 18, 19]. For the plasmonic SAM Skyrmion, we define $\chi$ as the angle of the interface projection of SAM, with respect to the x-axis [75, 76]. Clearly, $\chi$ has radial dependence within $r_2$, the Skyrmion boundary, and is plotted in Figure 3 as a function of distance $r$ from the vortex core, again for a coupling structure with $R = 2\lambda_{spp}$. We see that $\chi \sim 0$ for $r \to 0$ in eq. (3b), but away from the core, $\chi$ becomes negative and approaches -49° when $r \to r_2$. This variation of $\chi$ demonstrates that the plasmonic SAM Skyrmion, enclosed by the boundary $r_2$, deviates from a Néel type. Despite its $r$ variation, we define the helicity angle at the Skyrmion boundary ($r_2$) as a characteristic angle, which gives $\chi \sim -49°$. In addition, according to eq. (3b), a transition from a twisted to a Néel type Skyrmion is expected, as the size of the coupling structure increases, *i.e.* $R \to \infty$, because when $r \ll R$, then the $r$ dependent $S_\theta$ component approaches to zero. We also plot the structure size dependence of the characteristic Skyrmion helicity angle. A clear evolution of the helicity angle is observed from $\chi \sim -66°$ for $R = 1\lambda_{spp}$, to $\chi \sim -1°$ for $R = 100\lambda_{spp}$, confirming that a twisted Skyrmion transitions to a Néel type as the coupling structure size increases with respect to $\lambda_{SPP}$.

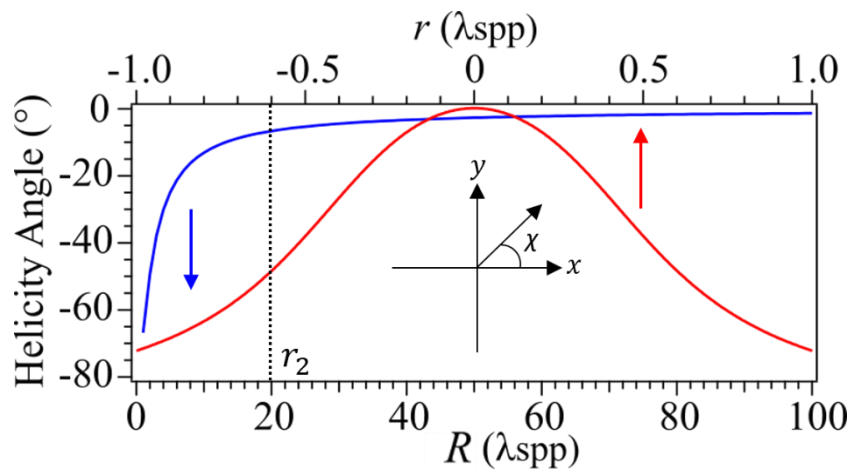

**Figure 3. Distance dependence of the helicity angle of the in-plane SAM (red), for coupling structure of $R = 2\lambda_{spp}$. The helicity angle approaches zero at the vortex core ($r$=0), and its magnitude increases with distance. The size dependence of the Skyrmion helicity angle taken at the Skyrmion boundary (blue), shows a clear transition from a twisted type for $R < 10\lambda_{spp}$ to a Néel type for $R > 10\lambda_{spp}$. Vertical arrows point to the corresponding axes of the curves, and the dashed line indicates $r = r_2$.**

## C. The SAM Skyrmion Number

To further evaluate the topological character of the plasmonic SAM texture, we calculate the Skyrmion density, $D$, from [12],

$$D = \frac{1}{4\pi} \boldsymbol{S_n} \cdot \left(\frac{\partial \boldsymbol{S_n}}{\partial x} \times \frac{\partial \boldsymbol{S_n}}{\partial y}\right) \qquad (2)$$

where $\boldsymbol{S_n}$ signifies the normalized SAM at each pixel of evaluation in the Cartesian coordinates. The calculated distribution of $D$ in the $x$ and $y$ coordinates for $L$=-1 vortex with $R = 2\lambda_{spp}$ is shown in Figure 4(a). Evidently, $D$ has a radial symmetry with the first maximum centered at $r_1$, which coincides with $S_{z,n} \sim 0$ in Figure 2(b). This is because the dominant contributions to $D$ represent regions where $\boldsymbol{S_n}$ is rapidly varying, which occurs where $S_{z,n}$ crosses zero. In the case of $L$=-1 vortex, $D$ is positive, because $\boldsymbol{S}$ changes from pointing up for $r < r_1$, to down for $r > r_1$. The Skyrmion number of the twisted SAM Skyrmion is obtained by integrating $D$ within the boundary enclosed by the circle of radius $r_2$, at which the SAM points purely in the negative $z$-direction. Performing the integration gives the Skyrmion number of $N$=1 for the $L$=-1 plasmonic vortex, where the integer value defines the quasiparticle as a Skyrmion [19]. Finally, because the sign of the SAM associated with an $L$=1 SPP vortex is reversed while its vorticity remains unchanged, its corresponding Skyrmion number would be $N$=-1, again associated with a Skyrmion.

We further show that the Skyrmion density distribution is intrinsically bounded by an L-line polarization singularity fringe of the surface electric fields. The L-line singularity [50] defines the interface where the in-plane electric fields are linearly polarized and the out-of-plane SAM polarization to passes through zero, i.e., $S_z = 0$. We numerically compute the L-line fringe distribution by calculating the ellipticity of the in-plane electric fields surrounding the vortex region, as shown in Figure 4(b). The minimum of the ellipticity of 1 corresponds to the circular in-plane polarization, which is known as a C-point singularity (the central point in Figure 4b) [50]. Away from the C-point, the SPP field becomes more elliptical and eventually passes through an L-line singularity at $r_1$. The ellipticity mapping identifies concentric ring fringes centered on the vortex core, with the inner most L-line, located at $r_1$, coinciding with the ring of dominant Skyrmion density. The spatial distribution of ellipticity is not limited by diffraction, and therefore can have substantially subwavelength dimensions, which can potentially be used to probe matter properties on a deep subwavelength scale [77].

The second L-line from the center marks where again $S_z = 0$ (dashed circle in Figure. 4), which encloses Skyrmion boundary. This $S_z = 0$ condition occurs, however, at a transition between where $S_z$ points directly down to directly up. Evidently here the L-line forms through a superposition of oppositely circulating in-plane fields on either side, that give the down/up $S_z$ orientation. This analysis shows that the L-line mapping provides a convenient method to locate both the dominant SAM Skyrmion density, as well as its boundary. The dominant SPP circular polarizations at the vortex core and their boundary cause the plasmonic Skyrmions to violate time-reversal symmetry.

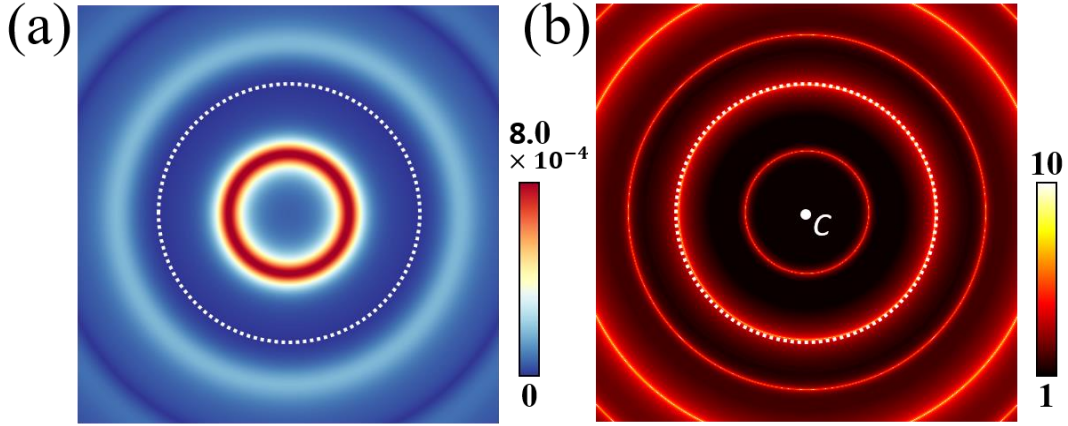

**Figure 4. (a) The plasmonic Skyrmion density *D* distribution that peaks at $r_1$ where $S_z = 0$. (b) A map of the L-line singularities, where the bright contrast shows concentric rings identifying regions where the in-plane plasmonic fields are dominantly linearly polarized. The inner ring occurs at $r_1$, where the Skyrmion density is maximum. The second ring marked by a dashed circle encloses the Skyrmion boundary at $r_2$. The C-point singularity (the point marked by C) is at the Skyrmion core.**

## D. Plasmonic SAM Hedgehog

One novel aspect of plasmonic Skyrmions is that the plasmonic fields are present on both sides of the interface. While the transverse cycloidal fields have opposite handedness across the interface because of the collective charge density oscillations, their in-plane field circulations are continuous due to the Maxwell's equation boundary conditions. Specifically, for an *L*=-1 vortex, $E_r$ at the core rotates in a counterclockwise manner (Figure S1), similar to left circularly polarized light. This handedness of field evolution must be consistent across the metal/vacuum interface, because the parallel SPP fields above and below the interface must be continuous according to their boundary conditions. In addition, because the existence of SPPs requires a sign change between the real parts of the dielectric functions, this causes the out-of-plane SAM, $\varepsilon(\boldsymbol{E}^* \times \boldsymbol{E})_z$, which is associated with the interface parallel electric fields, must have opposite sign. The in-plane SAM, however, which is locked to the counterclockwise *k$_{spp}$*-vector, remains parallel across the interface. As a result, the plasmonic SAM forms a Skyrmion texture corresponding to *N*=1 for a plasmonic vortex of *L*=-1, and the SAM texture is also imposed on the metal side forming an *N*=-1 Skyrmion, having an equivalent surface plane area. Thus, the SAM texture near the vortex core is composed of a Skyrmion pair with opposite topological charges, which together share similar texture to a spin hedgehog found in electronic band edges of gapped graphene and topological insulators, thus directly demonstrating breaking of the time-reversal symmetry [78, 79]. This is a particular property of plasmonic topological quasiparticles, which are defined by the properties of dielectric/metal interfaces, rather than the spin-spin interactions, as in the case of their magnetic homotopes.

Here we consider SAM within the primary ring ($r_1$), which for the Skyrmion pair forms a SAM hedgehog texture, where the total spin rotates by 2π on a sphere of radius $r_1$. In Figure 5(a), we plot the cross-section of the $\boldsymbol{S_n}$ texture, where the color scale shows the normalized SAM amplitude at each pixel, with the total distribution having no physical dimension), and in Figure 5(b) and 5(c), how the normalized SAM components change with *r* in the vacuum and metal sides. It is clear that the spin vectors start from pointing directly up in vacuum at the vortex core, and continuously rotate towards the interface plane upon approaching $r_1$; on the metal side, the dielectric function together with the field circulation causing the spin to point from directly down into the metal at the vortex core to in-plane in the same direction as the vacuum spin at $r_1$. Considering both sides of the interface, the complete 2π rotation with respect to the vortex core thus forms a 3D SAM hedgehog texture. We note that there is a SAM singularity at the core of a plasmonic vortex corresponding to the π phase jump in $S_{z,n}$ across the interface, and coincides with the C-point singularity. The phase jump is caused by the change in sign of the dielectric functions. The C-point singularity of a plasmonic Skyrmion creates the hedgehog spin distribution such as would be expected if it were created by a spin monopole.

Additionally, the rate of rotation of $S$ from vertical to horizontal in the metal, compared with the rate on the dielectric side, is sharper near the core, as seen in Figures 5(b)(c), but becomes sharper near $r_1$ on the vacuum side. The different behavior in the two media is a consequence of the superposition of the SAM contributed from the electric and magnetic fields circulations in the two media: $S_{z,n}$ from the in-plane electric field's circulation changes sign when the dielectric function passes through zero across the interface, but $S_{z,n}$ from the magnetic field's circulation remains the same, as seen in eq (3c). Consequently, on the dielectric side, the electric and magnetic contributions to the total SAM add up so that $S_{z,n}$ does not vary substantially within $r_1$. On the metal side, however, $S_{z,n}$ is determined by the differences between the electric and magnetic contributions, which causes its more rapid change as compared with the dielectric.

The coexisting pair of Skyrmions of opposite Skyrmion number occurs at quasi-2D metal/dielectric interfaces, which is a more tightly confined texture than the in-plane Skyrmion-anti-Skyrmion pairing in magnetic materials [80, 81]. Moreover, similar textures have been found in momentum space of photonic crystals, which can directly express the local Berry curvature [82].

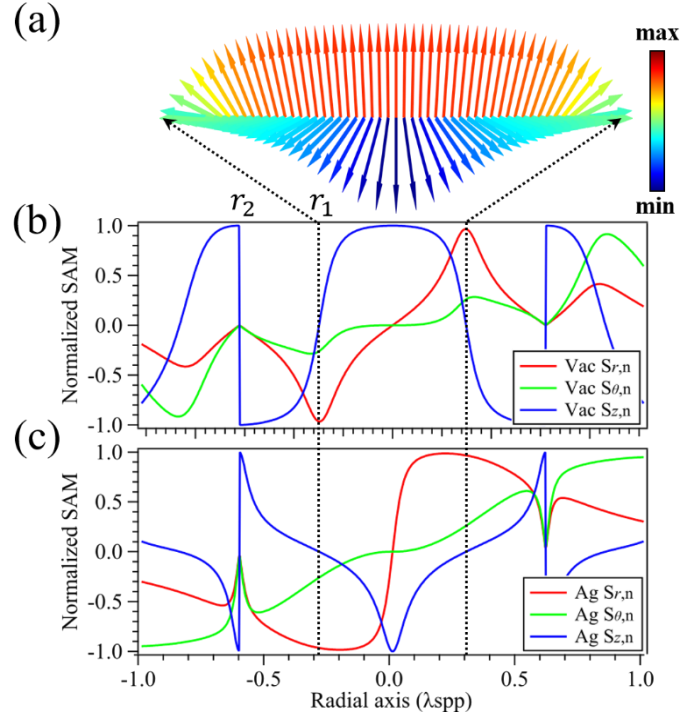

**Figure 5. (a) SAM hedgehog profile across the dielectric/metal interface. (b)(c) The radial changes of SAM components in the cylindrical coordinates, in the dielectric and metal sides, respectively.**

## III. Ultrafast Microscopy of Plasmonic Skyrmions

To validate the SAM Skyrmion existence and distribution experimentally, we perform ITR-2P-PEEM measurements that record a sequence of PEEM images of the plasmonic vortex at the center of a circular coupling structure when excited by circularly polarized light. ITR-2P-PEEM experiments primarily record the vectorial SPP field distributions at the vortex that excite the 2PP signal, but are not directly sensitive to SAM, which needs to be interpreted from the recorded 2PP signal distributions, which define the responsible fields and their associated spin textures. As shown schematically in Figure 6(a), we employ a pair of identical phase-delayed pump and probe pulses (20 fs, $\lambda_L = 550$ nm) that are normally incident onto the sample surface and are circularly polarized to launch SPP waves from the coupling structure. The SPP fields are generated with OAM $L=\pm 1$, depending on the SAM ($\sigma=\pm 1$) of light, and propagate towards the center, where they form a vortex. The coupling structure is composed of 3-fold concentric rings with the innermost having a radius of $5\lambda_{spp}$ [45].

Each slit is nominally formed ~100 nm wide and deep by focused ion beam etching of a polycrystalline Ag film. The coupling structure radii increase by $\lambda_{spp}$, so that the SPP fields that they individually generate add constructively, as in a grating structure, to intensify the coupled SPP field strength [83]. The primary data recorded in an ITR-2P-PEEM experiment are single pulse (delay τ independent) images as well as movies where the delay τ is advanced in ~100 as steps between the identical pump and probe pulses [40, 52, 53, 55, 57]. Retardation plates set the excitation laser polarizations.

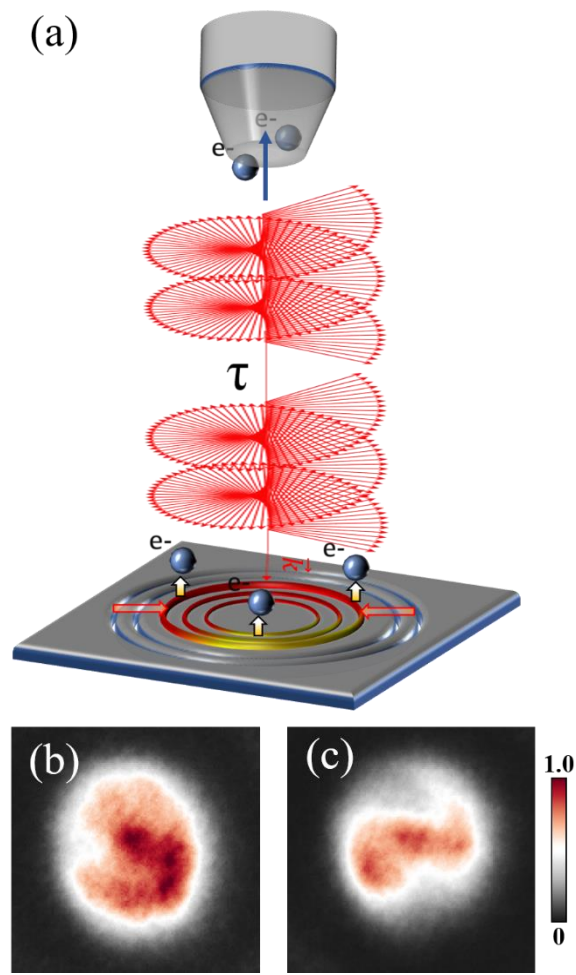

**Figure 6. (a) Schematic of the ITR-2P-PEEM experiment. (b)(c) Static, single pulse, PEEM images of an *L*=-1 and *L*=+1 plasmonic vortices. All experimental PEEM image presented have size of 2$\lambda_{spp}$.**

With identical, phase correlated pump and probe pulses we image the SPP fields. The signal has a contribution where the pump pulse generates the SPP field, and the probe pulse interferes with its longitudinal component. Thus, by advancing the pump-probe delay the interference creates a signal component that is spatially modulated with periodicity of the SPP wavelength; the interference pattern spatially propagates at the local speed of light as $\tau$ is advanced [54, 56]. The 2PP signal also includes a contribution from the transverse ($E_z$) component of the plasmon field, which does not contribute strongly to interference signal because it is orthogonal to the optical field [52, 55, 84]. Advancing the delay, thus, obtains detailed time-dependent imaging of fields as they form the vortex. The as-recorded, single pulse, photoemission images excited with the left and right circularly polarized light are shown in Figures 6(b)(c). Both PEEM images for the opposite helicity excitations show superficially similar circular distributions, *i.e.* within circular regions of diameter $\sim\lambda_{spp}$ with the strongest photoemission coming from the center, where $E_z\rightarrow0$, and therefore, the signal must have strong contributions from the in-plane fields. The complex mixture of the in- and out-of-plane field induced photoemission processes, however, complicates the SAM Skyrmion evaluation, because one cannot immediately separate its cylindrical components.

To analyze 2PP signals from the in-plane from the out-of-plane SPP fields, we Fourier filter the data from ITR-2P-PEEM experiments as follows [41]. The optical-SPP interference modulates ITR-2P-PEEM movies at the integer multiples (*n*=0,1,2,…) of the driving laser frequency, $n\omega_L$ [84]. To perform Fourier filtering of the signal, we Fourier transform (FT) the ITR-2P-PEEM movie, and then isolate one of its $n\omega_L$ components, to perform an inverse FT (IFT) [41, 85]. This filtering procedure eliminates the signal from the $E_z$ component of 2PP, which is not modulated by varying $\tau$ through the optical-SPP interference. Thus, in Figure 7, we obtain images by IFT of the *1*$\omega_L$ signal component that selectively shows the in-plane field evolution. The absolute value of the radial field amplitude, $|E_r|$, at $\tau \sim$19.1 fs ($\tau$ =0 fs delay is defined when the pump-and probe pulses overlap in time) is shown in Figure 7(a); it has the main contribution from the radial component $|E_r|$ of the SPP fields. One can see a clear focus of the radial field within the vortex core and a local minimum at the vortex ring ($r_1$), which match well with our analytical model in Figure 1(c).

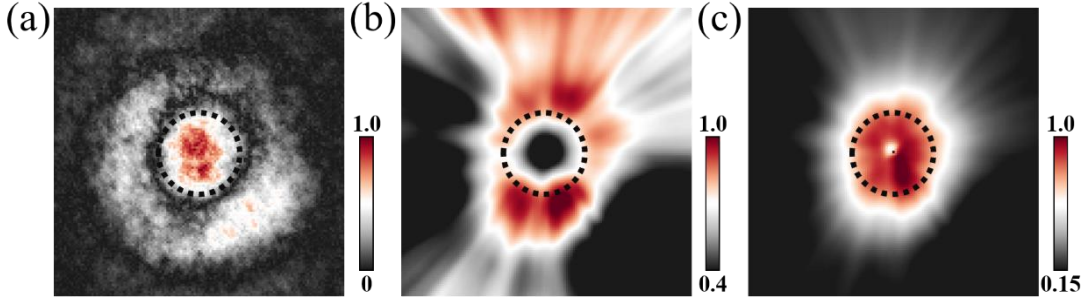

**Figure 7.** (a) $|E_r|$ component of the SPP vortex at τ~19.1 fs extracted by IFT of the ITR-2P-PEEM movie at $1\omega_L$. (b)(c) derived $|E_z|$ and $|E_\theta|$ components from (a). The image sizes are $2\lambda_{spp}$.

Because Maxwell's equations define the space and time evolution of the SPP fields at the sample surface, we can use the experimental $E_r$ distribution to derive the $E_z$ and $E_\theta$ components of the SPP fields [58]. The derived $|E_z|$ and $|E_\theta|$ distributions based on $E_r$ are plotted in Figures 7(b)(c). Because the signal processing involves spatial derivatives along the radial axis, the derived images are probably affected and exhibit related noise modulations in the tangential axis, as is evident from the diverging ray-like contrast in Figures 7(b)(c). This contrast, however, does not compromise evaluation of the vortex field distribution, because the major intensity regions, *i.e.* the primary vortex ring of $|E_z|$ and the slowly decaying $|E_\theta|$, in Figures 7(b)(c) agree with the corresponding analytical field distributions presented in Figures 1(b)(d). In addition, the dashed circle in Figure 7 marks the calculated location of the vortex ring, which confirm that our raw and derived field distributions follow the predicted ones from the Bessel functions in eqs. (2).

Finally, we can obtain the SAM component distributions from the field distributions algebraically based on eqs. (2) and (3). Because the radial component of the SAM in eq. (2a) follows $J_1^2(k_{spp}r)/r$, and the $E_z$ field component follows $J_1(k_{spp}r)$, therefore we can construct algebraically the experimental radial component of SAM based on the field component image in Figure 7(b). The derived distribution of $S_r$ is shown in Figure 8(a), from which it is evident that there exists a continuous change in the radial SAM. The strongest radial component is located near the vortex ring, which agrees with Figure 2(c). The z-component of the SAM can be obtained in a similar way, by using Figure 7(a) and (c), because $E_r \sim J_0(k_{spp}r) + J_2(k_{spp}r)$ and $E_\theta \sim J_0(k_{spp}r) - J_2(k_{spp}r)$, so their product defines the distribution of $S_z \sim J_0^2(k_{spp}r) - J_2^2(k_{spp}r)$, which is shown in Figure 8(b). To avoid confusion, our approach does not directly image of the SPP SAM distribution, but our analytical model of the SPP fields enables us to transform the field into spin distributions based on their analytical Bessel function profiles. Once a field component is determined experimentally by IFT analysis of an ITR-2P-PEEM sequence of images, the other field and SAM components follow mathematically from their Bessel function distributions, according to the analytical model. It is clear that $S_z$ evolves continuously from its positive maximum at the vortex core to a negative ring outside of the primary vortex field ring, and $S_r$ coincides with the vortex field ring, in agreement with the expected SAM component distributions from the analytical model in Figures 2(b)(c).

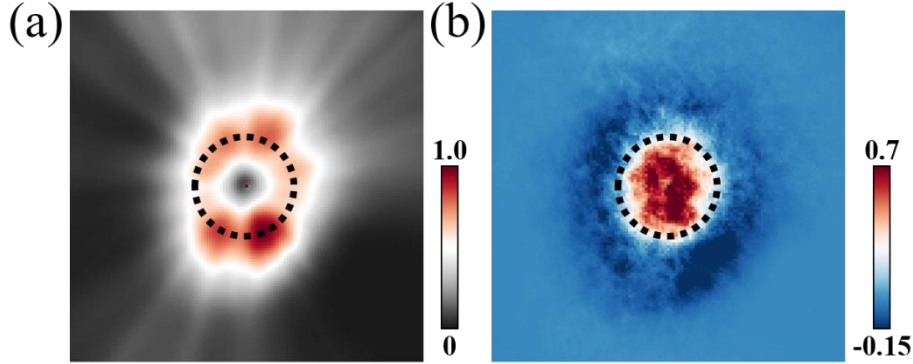

**Figure 8.** Derived $S_r$ and $S_z$ components of the plasmonic SAM, based on the electric field distributions in Figure 7.

## IV. Conclusions:

We have imaged and analytically described the vectorial components of a new quasiparticle, a twisted plasmonic SAM Skyrmion, at the core of a plasmonic vortex of OAM *L*=±1. Analogous to magnetic Skyrmion, the twisted SAM Skyrmion texture for L=-1 vortex has its spin pointing up at the quasiparticle center, and continuously reorients through horizontal at the vortex fringe to pointing down at its boundary. The quasiparticle boundary is defined by the second optical L-line singularity fringe, where the spin discontinuously changes from down to up. Within the quasiparticle, the in-plane spin gyration, characterized by the helicity angle, deviates the plasmonic SAM texture from the Néel type. Further analysis demonstrates that the twisted SAM texture approaches a Néel type for the radius $R_0 \gg \lambda_{spp}$. While we find a Skyrmion of Skyrmion number *N*=1 in the vacuum side for *L*=-1 vortex, the plasmonic SAM forms a mirror Skyrmion of charge *N*=-1 in the plasmonic metal, which together form a 3D Skyrmion pair across the interface. This analysis of plasmonic vortex and Skyrmion field distributions is an important conceptual advance because they are generated and evolve on the femtosecond time scale defined by duration of the excitation optical pulse, and they provide a fundamental understanding of the creation and annihilation dynamics for topological defects in optics, solid state systems, and in the context of cosmology. Such SAM texture could also be generated with stronger confinement in the lateral and vertical scales in multilayer metal dielectric films [86-88], thus enabling strong quasiparticle interaction between plasmonic SAM Skyrmion and other excitations such as conduction band electrons and excitons in semiconductors [19, 31, 32]. Finally, we imaged surface plasmonic field evolutions using ITR-2P-PEEM technique, and extracted the evolution of the radial electric field, which we use to derive the distributions for both the tangential and out-of-plane electric fields, and the distribution of the vectorial SAM components.

Our ultrafast imaging, together with the theoretical description, demonstrates the existence of a novel topological quasiparticle the plasmonic SAM Skyrmion. The SAM Skyrmion is a dynamical quasiparticle that is stable within the lifetime of SPP wave packet, during which it can potentially transfer its topological spin textures to other topologically trivial or nontrivial materials [89, 90], transferring spin and angular momentum to generate electron vortices [66], and Skyrmion number dependent chiral nano-object manipulations and sensing on ultrafast time scale [71, 91]. We have demonstrated that the fabricated geometrical charge of a topologically trivial material, can generate ultrafast topological spin textures that can be transferred and transiently modify the electronic properties of materials in the near field of surface plasmons, and that can even participate more intimately by acting as dielectric substrates for plasmonic metals. Finally, structuring SPPs in space can also create magnetic meron-like SAM textures [16], which can be generalized to meron-like lattices by SPP interference. Such quasiparticle lattice should possess periodic potential similar to optical lattices, thus it provides a broad platform for studying spin-dependent quasiparticle interactions, and atom or nanoparticle trapping [92, 93], and may find applications as dynamical nonlinear materials with similar functions as fabricated metamaterials or flat optics [94-96].

## Acknowledgements

This research was supported by the NSF Center for Chemical Innovation on Chemistry at the Space-Time Limit grant CHE-1414466. YD thanks R. Mong for consultation on the analytical theory.